# Reactivity of ultra-thin Kagome Metal FeSn towards Oxygen and Water

James Blyth[1,4,⊥], Sadhana Sridhar[1,⊥], Mengting Zhao[1,3,4,⊥,*], Sajid Ali[2,4], Thi Hai Yen Vu[1,4], Qile Li[1,4], Johnathon Maniatis[1], Grace Causer[1,4], Michael S. Fuhrer[1,4] Nikhil V. Medhekar[2,4], Anton Tadich[3,4], Mark Edmonds[1,4,5]

[1] School of Physics and Astronomy, Monash University, Clayton, VIC 3800 Australia

[2] School of Materials Science and Engineering, Monash University, Clayton, VIC 3800 Australia

[3] Australian Synchrotron, Clayton, VIC 3168 Australia

[4] ARC Centre for Future Low Energy Electronics Technologies, Monash University, Clayton, VIC 3800, Australia

[5] ANFF-VIC Technology Fellow, Melbourne Centre for Nanofabrication, Victorian Node of the Australian National Fabrication Facility, Clayton, VIC 3168, Australia

[⊥] These authors contributed equally to this work

\* Corresponding author: mengting.zhao@monash.edu

## Abstract

The kagome metal FeSn, consists of alternating layers of kagome-lattice $Fe_3Sn$ and honeycomb $Sn_2$, and exhibits great potential for applications in future low energy electronics and spintronics because of an ideal combination of novel topological phases and high-temperature magnetic ordering. Robust synthesis methods for ultra-thin FeSn films, as well as an understanding of their air stability is crucial for its development and long-term operation in future devices. In this work, we realize large area, sub-10 nm epitaxial FeSn thin films, and explore the oxidation process via synchrotron-based photoelectron spectroscopy using *in-situ* oxygen and water dosing, as well as *ex-situ* air exposure. Upon exposure to atmosphere the FeSn films are shown to be highly reactive, with a stable ~3 nm thick oxide layer forming at the surface within 10 minutes. Notably the surface Fe remains largely unoxidized when compared to Sn, which undergoes near-complete oxidation. This is further confirmed with controlled *in-situ* dosing of $O_2$ and $H_2O$ where only the $Sn_2$ (stanene) inter-layers within the

FeSn lattice oxidize, suggesting the Fe$_3$Sn kagome layers remain almost pristine. These results are in excellent agreement with first principles calculations, which show Fe-O bonds to the Fe$_3$Sn layer are energetically unfavorable, and furthermore, a large formation energy preference of 1.37 eV for Sn-O bonds in the stanene Sn$_2$ layer over Sn-O bonds in the kagome Fe$_3$Sn layer. The demonstration that oxidation only occurs within the stanene layers may provide new avenues in how to engineer, handle and prepare future kagome metal devices.

Introduction

The kagome lattice has attracted significant interest due to the coexistence of a topologically non-trivial band structure and strongly correlated phenomena that arise from the unique geometry of corner-sharing triangles[1-4]. In the absence of electron-electron interactions and spin-orbit coupling (SOC), the two-dimensional (2D) kagome lattice hosts two intersecting Dirac bands and a single flat band. Introducing SOC and correlations due to electron-electron interactions is thought to result in magnetic ordering, opening a gap at the Dirac point and enable dissipationless transport of charge along a 1D chiral edge channel via the quantum anomalous hall effect[5-9]. However, with increasing number of stacked kagome layers, the bandgap may diminish, disappearing at a critical thickness as the system evolves into a 3D magnetic Weyl semimetal[10]. Hence this transition has sparked an interest in achieving ultra-thin kagome metals. The flat band in the kagome lattice has the potential to induce exotic electronic phases such as skyrmion superconductivity[11] and even the fractional quantum Hall state[8]. All these exciting properties contained within a single system make kagome metals an ideal platform both for fundamental physics and for promising future technologies including topological quantum computing, dissipationless electronics, and spintronics.

Among the kagome metal families, FeSn stands out for its above room-temperature (~365 K[12, 13]) anti-ferromagnetism and quasi-2D structure as shown in Fig.1a, where the kagome layer (Fe$_3$Sn) and the stanene layer (Sn$_2$) alternatively stack along the c-axis. This is quite different to Fe$_3$Sn$_2$ and Fe$_3$Sn, where the stanene inter-layers physically and electronically separate the kagome layers, potentially reducing the need to reach the 2D regime for FeSn to exhibit a bandgap. Rich physical phenomena have been found in bulk and thin-film FeSn, including Dirac fermions[3, 12], giant periodic pseudomagnetic fields[14], symmetry-breaking electronic orders[15], and signatures of the flat band[3, 16, 17], showing great potential for applications in spintronics. However, to date only FeSn films with thicknesses greater than 10 nm have been achieved, and a strategy to realize isolated Fe$_3$Sn kagome layers remains unsolved.

Currently, most research on kagome metals focuses on exploration of magnetic and topological properties, and is generally achieved by minimizing the exposure to atmosphere with careful preparation in inert environments or ultrahigh-vacuum conditions, followed by encapsulation with air-stable overlayers such as carbon or amorphous BaF$_2$[18, 19]. Yet, little attention has been paid to the stability of kagome materials in ambient, atmospheric conditions and whether an oxide surface layer will form in a similar manner to the topological magnetic insulator MnBi$_2$Te$_4$[20], reconstruct the surface in the case of the topological insulator Bi$_2$Se$_3$[21] or degrade like black phosphorus[22]. A comprehensive understanding of the oxidation process of kagome metals under ambient conditions is crucial before they can be utilized for future next-generation electronics applications.

In this study, we realize high-quality ultra-thin kagome metal FeSn films grown via molecular beam epitaxy (MBE) on Si (111). The stability of the sub 10 nm-thick FeSn films under various oxygen-containing gaseous environments was studied using synchrotron-based photoelectron spectroscopy (XPS). It was found that the FeSn films were highly reactive to atmosphere, achieving a passivated oxide state within 10 mins. Surprisingly, studies under controlled doses of pure O$_2$ and H$_2$O revealed that the Fe and Sn in the kagome layers barely oxidize, whilst Sn in the stanene layers near the surface approached complete oxidation. This is well supported by first principles density functional theory (DFT) calculations, which found the formation energy of Sn-O bonds in the stanene layer significantly smaller than Sn-O bonds in the kagome layer, while the Fe-O bond could not stably form. This suggests the kagome layers remain largely unaffected even when the stanene layers completely oxidize, and that is in excellent agreement with our *in-situ* exposure results.

## Results and Discussion

FeSn thin films were successfully grown on Si (111) substrate via MBE (see Methods for details). Figure 1(b) depicts the diffraction pattern of the film measured via reflection high-energy electron diffraction (RHEED) where the uniformity and sharpness of the streaks indicate a single-crystal, high-quality FeSn film. The 6-fold crystal structure and large area crystallinity was verified by using low-energy electron diffraction (LEED), as shown in Figure 1(c). Large-area topographic information of the sample surface was obtained by atomic force microscopy (AFM) measurements. As demonstrated in Fig. 1(d), high-coverage, uniform, ~8 nm thick FeSn films were successfully synthesized on the Si substrate. AFM images were also taken on the surface of 5 nm and 9 nm thick FeSn films, with high (>85%) coverage and similar quality (Fig. S1). Figure 1(e) shows XPS data of pristine FeSn taken at $hv = 880$ eV, with the Fe $3p$ and Sn $4d$ core levels observed in the correct chemical stoichiometry, Fe:Sn = 1:1, after factoring in the respective different photoionization cross-sections. Alongside the XPS, secondary electron cut-off (SECO) measurements were performed, where the extracted work function (see Fig. S2) for pristine FeSn was $4.1 \pm 0.1$ eV. This is comparable to metals commonly used as electrical contacts (Au: 5.06 eV[23], Pt: 5.54 eV[23] and Al: 4.04 eV[23]) and would help with device fabrication and achieving ohmic contact. To confirm the electronic bandstructure, angle-resolved photoelectron spectroscopy (ARPES) was performed as shown in Fig. 1(f). The spectra show the clear presence of a pair of Dirac bands intersecting at the K points. The broadness of the bands is attributed to the nature of thin-film measurements, sample quality and possible band splitting due strong spin-orbit coupling that is not captured due to the detector resolution. The Dirac bands were fitted using the massless Dirac equation, allowing us to extract the Fermi velocity of $v_F = 1.5 \pm 0.3 \times 10^5 \ m/s$ along the ΓK direction, which is within uncertainties of the reported bulk velocity[3] of $1.7 \pm 0.2 \times 10^5 \ m/s$. The Dirac point was subsequently determined using the fitted bands to be, $E_D \sim 70 \ meV$ below the Fermi level, which is a considerably different to bulk results[4,12] of $\sim 0.4 \ eV$ below Fermi and other thin films[14,15,18] of $\sim 0.36 \ eV$. This indicates a significant doping shift that may be caused by the ultra-thin nature of the films and the highly n-doped Si substrate used.

To understand the reactivity of FeSn to oxygen, *in-situ* controlled levels of oxygen exposure were performed and studied using XPS. Figure 2(a) shows the evolution of the Fe $3p$ core level spectra, taken at $hv = 880$ eV, with increasing oxygen coverage (expressed in Langmuir). Based on the lattice structure, Fe is initially found in a single chemical environment (aside from a small oxide component discussed below), and this is confirmed by the single Fe $3p$ peak measured in the pristine case (in orange). The Fe $3p$ peak is described by a highly asymmetric Voigt distribution at binding energy $53.8 \pm 0.1$ eV, where the Fe $3p_{1/2}$ and Fe $3p_{3/2}$ spin-orbit splitting is not clearly distinguishable (<1 eV) and hence has been fitted as one peak. After the initial dose of $2.45 \times 10^3$ L, two oxide peaks $Fe^{2+}$ (dark blue) at $54.6 \pm 0.1$ eV and $Fe^{3+}$ (light blue) at $55.8 \pm 0.3$ eV, emerged with a combined oxide-to-main peak ratio of $7.26 \times 10^{-4}$. The ratio steadily increases and begins to plateau at $1.59 \times 10^{-3}$ for the last exposure of $3.83 \times 10^5$ L, indicating that less than 0.1% of Fe is oxidizing.

In contrast, the evolution of the Sn $3d$ spectra under the same oxygen exposure (Fig. 2(b)) taken at $hv = 1340$ eV, suggests that tin readily undergoes oxidation. The pristine Sn $3d$ spectrum has been fitted using three distinct environments: A Sn-stanene environment (orange), Sn-kagome environment (green) and Sn-oxide (blue), each having a fixed Sn $3d_{3/2}$, $3d_{5/2}$ spin-orbit splitting of 8.4 eV. The Sn-stanene doublet is positioned at 486.2 eV for the $3d_{5/2}$ peak (Sn $3d_{5/2}$ in Sn-metal is at 484.9 eV[24]) and is accompanied by broad satellite peaks, while the Sn-oxide and Sn-kagome doublets are at 486.5 – 487.2[24] eV and ~489 eV, respectively. The large difference in the intensity of signal from the Sn-stanene doublet and Sn-kagome doublet alludes to the terminating layer of the film being dominated by stanene, as XPS is extremely surface sensitive. Though the pristine film should not have any oxide present, the oxide doublet was included to account for any contamination during sample transfer and does indicate some exposure of oxygen occurred (also can be seen in the Fe $3p$ spectra in Fig 2(a)). The Sn is shown to oxidize immediately after the first oxygen dose, with the oxidation doublet clearly growing with each subsequent oxygen dosing. Saturation begins to occur at higher oxygen coverages of $1.22 \times 10^5$ L, where the growth rate of the oxide peak slows but does not reach zero, with the final coverage of $3.83 \times 10^5$ L resulting in an oxide-to-pristine peak ratio of 1.09. A more subtle observation is the persistence of the Sn-kagome doublet between each dose (see close inspection of spectra in Fig. S3). Along with the clear relative reduction in the pristine Sn-stanene doublet area (in orange), it appears the oxidation of Sn is heavily concentrated to the Sn in the stanene layers versus the kagome layers.

Analyzing the various oxide species of Fe and Sn and comparing their relative abundances is challenging when using only the Fe and Sn core levels, especially due to the close binding energies of SnO (486.5 eV)[25] and $SnO_2$ (486.7 eV)[25] in the Sn $3d$ spectra. Therefore, we shift our focus to the O1s core level to more accurately characterize the oxide species that have formed. As shown in Fig. 3(a), the O1s peak consists of three distinct oxide species, located at approximately 532 eV, 531 eV and 530 eV, corresponding to $SnO_2$[26], SnO[26] and the mixture of Fe oxides, respectively. The Sn oxides (purple and pink) dominate over the slower-forming Fe oxides (green), with the Fe oxides accounting for only 5% of all oxides formed at the final coverage of $3.83 \times 10^5$ L. This further confirms that Sn oxidizes over Fe, where the relative magnitude is only feasible if the oxidation is concentrated to the Stanene layers. In addition, the sample predominantly consists of the more chemically stable $SnO_2$ over SnO, where the multi-step oxidation of Sn, Sn→$Sn^{2+}$→$Sn^{4+}$, likely drives the passivation of the film. Based on the Fe, Sn and O core levels, Sn is clearly more susceptible to oxidation compared to Fe, with $SnO_2$ being responsible for the film passivation. Further, the lack of chemical change to the Fe core level and preservation of the Sn-kagome doublet signal in the Sn core level, asserts that the oxidation of Sn must be heavily limited to the stanene layers, leaving the kagome layers largely unaffected.

To explain why Sn in the stanene layer is the most susceptible to oxidation in FeSn films, we provide a qualitative argument from its lattice structure. Consider the crystal structure of FeSn, as shown in Fig. 1 (a). The termination of the FeSn film can either be the stanene layer or the kagome layer. The kagome layer features six Fe-Sn bonds with each Sn atom encircled by the Fe atoms, resulting in robust Fe-Sn bonding and no inter-layer bonding. This arrangement creates a relatively enclosed structure for both Fe and Sn, with attempts at bonding oxygen requiring breaking a considerable number of bonds. On the other hand, in the stanene layer, Sn atoms are bonded to three other Sn atoms and there are three interlayer Fe-Sn bonds. This arrangement introduces more Sn atoms with accessible surfaces and potential binding sites for oxygen atoms, increasing the likelihood of oxygen interactions. The susceptibility is possibly enhanced by the Sn-Sn bonds being longer in length (approximately 3.06 Å) versus the Fe-Sn bonds (ranging from 2.65 Å to 2.73 Å). To fully understand the oxidation process, Density Functional Theory (DFT) calculations (described in Methods) were performed to ascertain the most chemically stable oxide. Firstly, attempts to bond oxygen to Fe with a kagome terminated surface failed, resulting in a breakdown of the simulation, thus affirming its strong resistance to oxidation. Next, the formation energy of Sn-O with the stanene termination and versus the kagome termination was calculated, finding the stanene layer was significantly preferred by 1.37 eV. This strongly agrees with experimental results and reinforces the argument that the stanene layer is considerably more prone to oxidation than the kagome layer. As the kagome layers remain mostly pristine, with the stanene layers oxidize, it may be speculated that the $Fe_3Sn$ layers become physically de-coupled as shown in the schematic in Fig. 3(b).

Now we turn to understanding whether the oxidation of the FeSn film occurs only at the surface, or penetrates throughout the material, by performing depth-dependent XPS. The effective probing depth ($d$) of XPS measurements is dominated by the mean-free path ($\lambda$) of the photoelectrons, which can be estimated by using the material lattice constant ($a$) and the photoelectron kinetic energy ($E_K$) via the following relation $d \sim \lambda = 0.41 a^{3/2} \sqrt{E_K}$, as depicted in Fig. 4(a). Hence, by increasing the incident photon energy, the photoelectron kinetic energy increases, and the effective probing depth increases. To ensure the probing depths of the Fe $3p$ and Sn $3d$ core levels were the same, photon energies (Fe $3p$: 160 eV, 340 eV, 880 eV, Sn $3d$: 620 eV, 800 eV, 1340 eV) were chosen to ensure approximately the same kinetic energies of $E_K \sim 100, 280$ and $820 \, eV$, with corresponding probing depths of 0.6, 1 and 2 nm. Figure 4(b) and (c) compare the Fe 3p and Sn 3d spectra taken from the $1.22 \times 10^5$ L oxygen exposure corresponding to the three probing depths. Using the same color coding from Figure 2, the orange peaks represent pristine chemical environment, blue represents oxide environments and the green doublet in the Sn 3d signal from Sn in the kagome layers. While it has been earlier discussed, Fe remains largely un-oxidized, and comparing the Fe core level (Fig. 4b) between the 160 eV, 340 eV and 800 eV spectra, the Fe oxide species that are observed are clearly more concentrated at the surface.

The most surface sensitive Sn 3d spectrum (hv = 620 eV) has a large oxide-to-peak ratio of 1.67, which is considerably higher than the 800 eV and 1340 eV spectra of 1.22 and 0.77 respectively. This convincingly demonstrates the Sn oxide is confined to the surface of films. Another observation made from the three photon energies is the relative increase in the Sn-kagome doublet area (green) as the photon energy increases in Fig. 4(c). This is expected, as with greater probing depths, more kagome layers are captured and the signal from the stanene-terminating surface is less significant.

Confirming the oxide is confined at the surface, allows us to estimate the oxide thickness using the following overlayer attenuation relation[28], $d_{oxide} = \lambda \cdot Sin(\theta) \cdot \ln\left(\left(\left(I_{oxide}^{exp}/I_{element}^{exp}\right)/\beta\right) + 1\right)$, where $\theta$ is the angle between the surface plane and the electron detector, $\beta = I_{oxide}/I_{element}$ is the intensity area ratio between infinitely thick element oxide and element, and $I_x$ is the peak intensity area of the respective species. Additionally, as primarily Sn oxidizes, we only focus on the Sn 3d spectra and enhance the accuracy of the measurement, by taking the average thickness between the 800 eV and 1340 eV Sn 3d spectra. Figure 5(a) re-plots the evolution of the Sn 3d core level with increasing oxygen coverage and Fig. 5(b) plots the corresponding oxide thickness verses coverage. As previously mentioned, the pristine Sn 3d spectrum contained a measurable amount of Sn oxide that equates to a thickness of $0.5 \pm 0.1$ Å. At relatively low doses, the oxide thickness rapidly increases to $5.0 \pm 0.2$ Å with a coverage of $1.36 \times 10^4$ L, with the growth rate decreasing with more oxygen exposure. A final oxide thickness of $12 \pm 3$ Å was calculated at an $O_2$ exposure of $3.83 \times 10^5$ L, but is not considered the saturation point. The saturation point requires the growth rate to reach zero, where a stable, maximal oxide thickness has formed and needs further study with higher oxidation doses to ascertain.

The kagome metal FeSn films reactivity to water was also studied via *in-situ* gaseous water exposure experiments. Figure. 5(c) shows the evolution of the Sn 3d core level under a series of exposures, and the oxidation trend is similar to that seen for oxygen exposure (Fig. 5(a)), indicating a similar oxidation mechanism (also see Fig. S4 for Fe 3*p* and O 1*s* spectra). Careful comparison reveals that while the peak intensity of the Sn-oxide for the oxygen exposure is greater than that for water, the water Sn-oxide has a larger width and hence the total relative areas are near-identical. This follows with the oxide thickness (Fig. 5(d)) being in parallel to the $O_2$ exposure in shape and magnitude, reaching a thickness of $13 \pm 3$ Å at the final exposure of $4.08 \times 10^5$ L. The FeSn film exposed to air represents a marked contrast to $O_2$ and $H_2O$ exposures, serving as the fully saturated/oxidized case. This assertion primarily arises from the many orders of magnitude higher pressure ($\sim 10^{-5}$ mbar vs $\sim 10^3$ mbar) pressure that the sample is subjected to, resulting in significantly higher doses of $O_2$ and $H_2O$, and other atmospheric species. As shown in Fig. 5(e), within 1 minute of air exposure, the majority of Sn has been oxidized with an oxide-to-main peak ratio of 2.63, highlighting that FeSn is highly reactive in atmospheric conditions. With continued exposure, the oxide-to-main peak ratio between 1 hour, 12 hours, and 70 hours does not significantly change, indicating that a stable saturation point has been reached. Quantitatively, Fig. 5(f) plots the calculated oxide thickness, where the saturated oxide thickness is $33 \pm 2$ Å, varying marginally by 1 Å for the last three exposures. The comparisons between the three exposure times and the subsequent calculated oxide thickness demonstrates the reactive nature of FeSn films but also indicates the stability of the film to remain intact and passive.

## Conclusion

In this work, we successfully demonstrated large-area, sub-10 nm FeSn films, epitaxially grown on Si (111) via MBE. The films were exposed to *in-situ* dosing of $O_2$ and $H_2O$ and *ex-situ* air exposure and studied via synchrotron-based photo-emission spectroscopy. Within 10 minutes of air exposure, a ~3 nm thick oxide layer formed, which passivates the remaining film. Controlled *in-situ* dosing of $O_2$ and $H_2O$ revealed Fe to be extremely resistant to oxidation, while Sn was found to readily form both SnO and $SnO_2$. The lack of Fe oxide and the persistence of the Sn in the kagome layer suggests that only Sn in the stanene layers oxidizes, leaving the $Fe_3Sn$ kagome layers largely unaffected. These results are in strong agreement with DFT calculations. Provided just the stanene layers oxidize, oxygen intercalation may be used to break the inter-layer bonding and increase physical spacing between the kagome layers (as seen in Fig. 3(b)), possibly electronically and physically decoupling the kagome layers. This is similar to the situation seen in ionic intercalation of transition metal chalcogenides such as $NbSe_2$[29] and $PdTe_2$[30], where intercalation decouples the interaction between adjacent $NbSe_2$ layers, making the intercalated bulk exhibit the same electronic structure as monolayer NbSe$_2$ and even tailoring the superconductivity. Our results not only provide insights of the reactivity of FeSn to O$_2$ /H$_2$O exposure, but more importantly, may offer a potentially viable process for realizing decoupled Fe$_3$Sn kagome layers via O$_2$ /H$_2$O dosing otherwise not achieved via conventional methods.

## Methods

### Growth of FeSn thin films via molecular beam epitaxy.

The FeSn films were grown on *p*-type Si (111) substrates that were flash annealed *in-situ* to form the standard (7x7) surface reconstruction. The evaporation temperature for Fe (Sigma-Aldrich 99.98%) was 1300 ˚C using an MBE Komponenten HTEZ effusion cell and 1020 ˚C for Sn (Sigma-Aldrich 99.999%) housed in an MBE Komponenten dual source effusion cell. The evaporation rates were 1:1, and were calibrated using a quartz crystal microbalance (QCM). During deposition, the Si (111) substrate was heated to 350 – 370 ˚C, with an initial thin Sn buffer layer deposited to reduce the effects of islanding and enhance the formation of a commensurate and epitaxial film. Film quality was monitored during deposition using reflection high-energy electron diffraction (RHEED) using a beam voltage and current of 15 kV and 1.5 A respectively.

### Synchrotron-based photoelectron spectroscopy including angle-resolved photoelectron spectroscopy (ARPES) and X-ray photoelectron spectroscopy (XPS).

As-grown FeSn thin-film samples were transported from the MBE chamber at Monash University to the Soft X-ray Beamline at the Australian Synchrotron using an ultra-high vacuum (UHV) suitcase to maintain their pristine nature. Low-energy electron diffraction (LEED) was performed prior to the photoelectron spectroscopy measurements.

ARPES measurements were conducted on the Toroidal Analyzer endstation at T = 80 K using photon energies between 80-120 eV. The total convolved energy resolution was 100 meV. XPS measurements were taken on the PREVAC endstation with a SPECS Phoibos 150 analyzer. The total energy resolution of the system was approximately 50 meV. For probing oxidation depth of the films, the Fe 3*p* core level was measured at photon energies of 160 eV, 340 eV, 880 eV, whilst the Sn 3*d* and O 1s core levels were measured with photon energies of 600 eV, 800 eV, 1340 eV. All photon energies were calibrated by measuring the binding energy of Au

4f core level or Fermi edge of a gold foil reference in electrical contact with the sample. The work function was extracted using secondary electron cutoff spectroscopy (SECO) using a -9 V bias at 100 eV photon energy at a resolution of 20 meV.

*In-situ* oxygen/water exposure experiments were performed in the preparation chamber of the PREVAC system with base vacuum of ~$10^{-9}$ mbar. Pure oxygen and water species were introduced into the preparation chamber via precision leak valves to a controlled partial pressure of ~$1 \times 10^{-5}$ mbar. The dosage/coverage of the surface was calculated in Langmuir for the $O_2$ and $H_2O$ exposure experiments (1 Langmuir equates to exposure to a gas at $10^{-6}$ Torr for one second). Air exposure experiments were done via venting the sample in the load-lock chamber of the Pre-vac endstation by nitrogen gas and then exposing to atmosphere. XPS peak fitting was done in IGOR using a Shirley background subtraction and Voigt distributions.

## Atomic force microscopy (AFM).

Overall film morphology and growth quality across large areas was confirmed with atomic force microscopy. In order to minimize air exposure prior to AFM measurements, as-grown FeSn thin-films were directly transferred from the MBE system to an interconnected argon atmosphere glovebox, then vacuum sealed and transferred to the Melbourne Centre of Nanofabrication (MCN) cleanroom facility. AFM measurements were performed within 5 min of exposure to atmosphere. AFM measurements were conducted on a Bruker AFM system, with silicon tips in contact mode.

## Density functional theory (DFT) calculations.

We performed first pricniples density functional theory calculations to compute the adsorption energies of oxygen 'O' atoms on FeSn. For calculation of total energy and ground state geometry we used version 5.4.4 of the Vienna Ab Initio Simulation Package (VASP)[31, 32]. For accurate calculation of electron spin density close to the nuclei, the projector augmented wave method (PAW)[33] was applied together with a plane wave basis set- Pristine FeSn was first geometrically optimized using the conventional cell and a 9×9×9 Monkhurst-Pack reciprocal space grid. Bulk FeSn has hexagonal (space group P6/mmm) crystal structure, comprising of 10 kagome nets of 'Fe' and 'Sn', separated by 'Sn' buffer layers. 2x2x2 supercells of FeSn (periodic in lateral direction and truncated along vertical direction), containing 48 atoms in total, were then realized. A large vacuum region of 20 Å width was used to ensure no interaction of the truncated FeSn supercell with its periodic images along the vertical direction.

All the supercells were allowed to fully relax using a plane wave cut-off of 700 eV for a maximum force of 0.001 eVÅ-1. K-point convergence was checked and finally a k-point mesh of 3x3x1 was used for all calculations. Attempts to adsorb a single 'O' atom with Fe resulted in the destruction of the super cell in the simulation. Adsorption of single 'O' atom to Sn was found to be energetically favorable on the 'Sn' buffer layer compared to the 11 kagome layer by 1.37 eV. Calculations with different numbers of 'O' atoms adsorbed on the Sn buffer layer were subsequently performed. Our calculations reveal that FeSn-5O i.e. 5 'O' adsorbed on 'Sn' buffer layer (which has 8 Sn atoms) is the most energetically favorable system.

## Acknowledgements

J.B., M.Z., S.A., G.L.C, M.S.F., N.V.M., A.T. and M.T.E. acknowledge funding support from ARC Centre for Future Low Energy Electronics Technologies (FLEET) CE170100039. M.T.E. acknowledges funding support from ARC Future Fellowship FT2201000290. M.Z., A.T., M.T.E. acknowledge funding support from ANSTO-Postdoc Fellowship. M.Z. acknowledges funding support from AINSE Early Career Research Grant. J.B. acknowledges this research was supported by an AINSE Ltd. Postgraduate Research Award (PGRA). Part of this research was undertaken on the soft X-ray beamline at the Australian Synchrotron, part of ANSTO. This work was performed in part at the Melbourne Centre for Nanofabrication (MCN), the Victorian Node of the Australian National Fabrication Facility (ANFF). S.A. and N.V.M. gratefully acknowledge the computational support from the Pawsey Supercomputing Facility and the National Computing Infrastructure funded by the Australian Government.

Figures

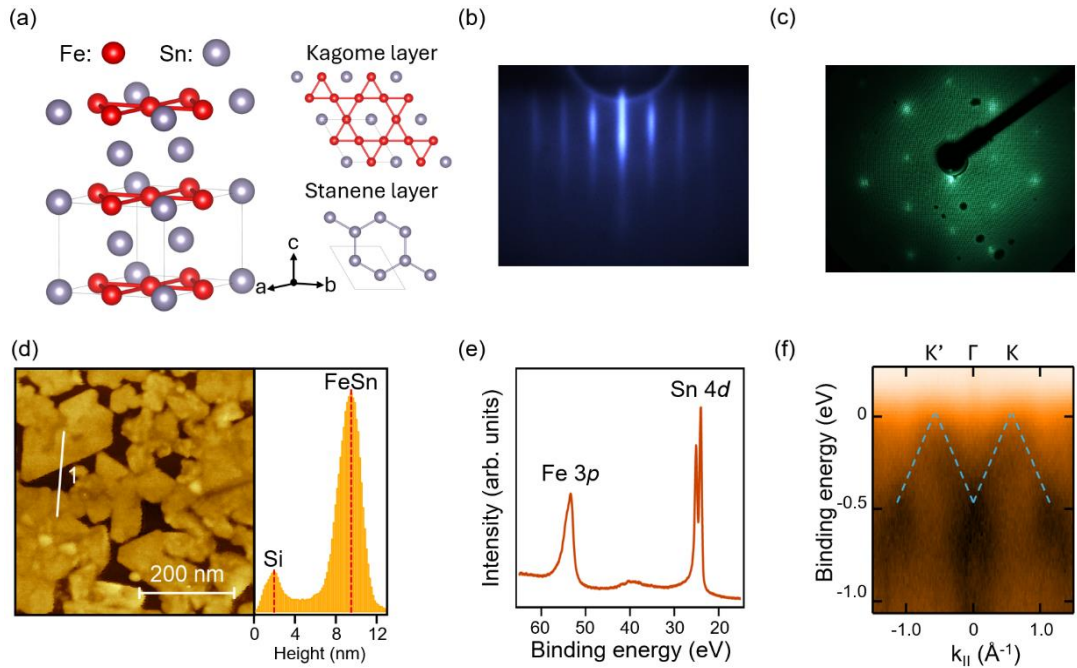

Figure 1. Growth and characterization of FeSn films. (a) Schematic of the FeSn crystal structure (from Materials project[27]). The left panel shows a 3D schematic of FeSn atomic structure which has a space group P6/mmm (191), and lattice constants a = b = 5.3 Å and c = 4.5 Å. The right panel shows the in-plane schematic of the two layers that form the FeSn lattice in an alternating stacking sequence: the $Fe_3Sn$ kagome layer and $Sn_2$ honeycomb Stanene layer. (b) Reflection high-energy electron diffraction (RHEED) pattern (15 kV, 1.5 A) of an as-grown FeSn film on the Si(111) substrate. (c) Low-energy electron diffraction (LEED) pattern of pristine FeSn film, taken at 73 eV capturing the first Brillouin zone. The symmetry is 6-fold triangular, with sharp spot sizes indicating single crystal growth. (d) Left panel: 500 nm x 500 nm atomic force microscopy (AFM) image of FeSn film. Right panel: Height profile histogram showing the average film thickness is approximately 8 nm. (e) X-ray photo-emission spectroscopy (XPS) taken at 880 eV on a pristine FeSn film, capturing the Fe 3p and Sn 4d core levels. The relative area ratio verifies the 1:1 stoichiometry. (f) Angle-resolved photo-emission spectroscopy (ARPES) spectra taken at $hv$=104 eV along the ΓK high-symmetry direction. The dotted blue lines are linear fits to the pair of Dirac cones observed in thin FeSn film sample.

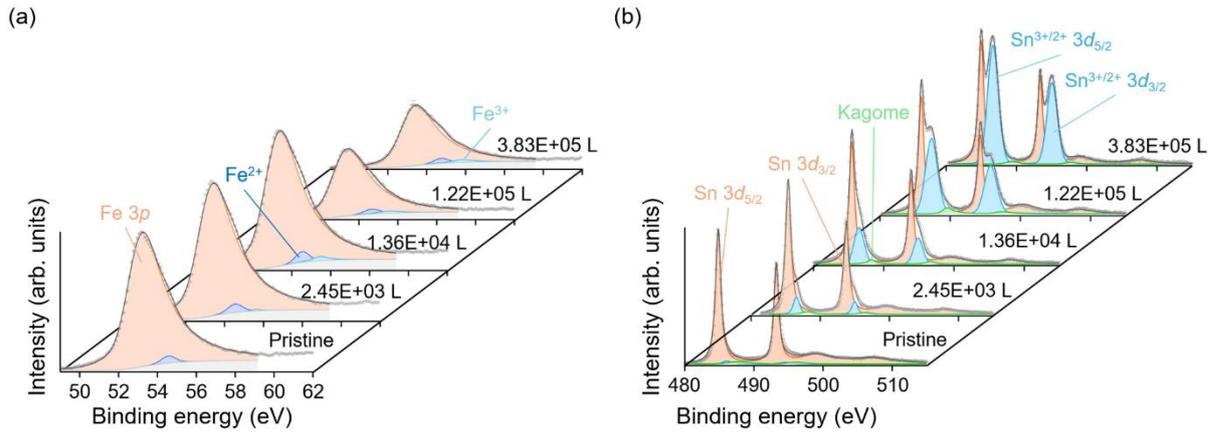

Figure 2. Evolution of the ultra-thin FeSn film after exposure to $O_2$ expressed in Langmuir (L). (a) Fe 3p core level and (b) Sn 3d core level taken at $h\nu$ = 880 eV and 1340 eV, respectively. The peaks corresponding to the (a) Fe environment (b) Sn environment associated with pristine FeSn are shown in orange. The oxide peaks are depicted in blue where two distinct Fe oxide species ($Fe^{2+}$ and $Fe^{3+}$) can be distinguished but $Sn^{2+}$ and $Sn^{4+}$ have been fitted as one. The green doublet seen in the Sn 3d spectra originates from the Sn in Kagome surface-terminating regions.

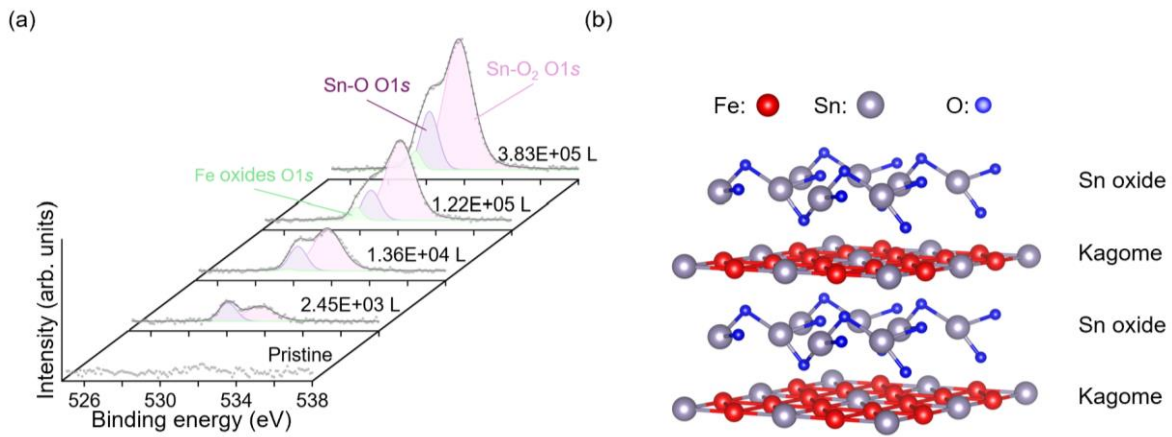

Figure 3. Evolution of oxide species present in the FeSn ultra-thin film after exposure to $O_2$ expressed in Langmuir (L) and proposed oxidized structure. (a) shows the evolution of the O 1s core level spectra with increasing coverage taken at a photon energy of 1340 eV. The spectra are fitted with three peaks corresponding to $Sn-O_2$ (pink), Sn-O (purple) and Fe oxides (green). The relative areas of these oxide peaks indicate a clear preferential oxidation of Sn over Fe. (b)

Schematic of the proposed oxidation in the FeSn structure showing the preferential oxygen bonding to the stanene layer over the kagome layer.

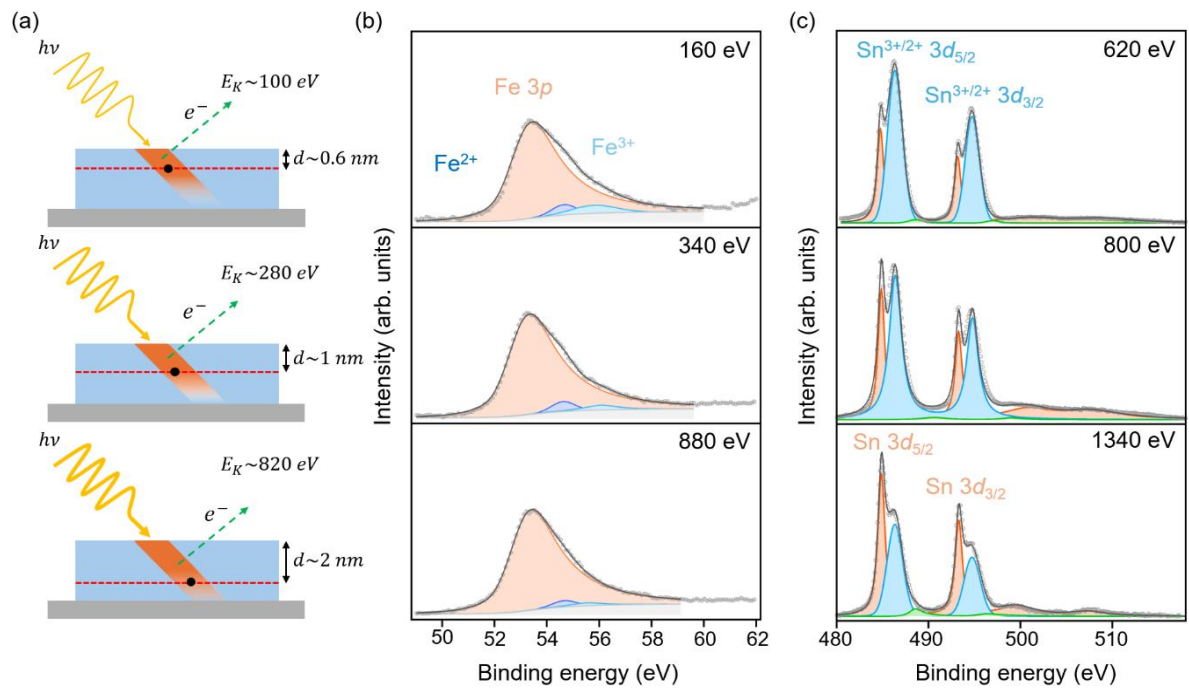

Figure 4. Illustration of depth-dependent XPS and spectroscopy of Fe 3p and Sn 3d core levels. (a) Schematic illustrating the change in effective probing depth with increasing photon energy calculated as described in main text. (b) and (c) are the depth-dependent photoelectron spectroscopy of Fe 3p and Sn 3d core levels after $1.22 \times 10^5$ L oxygen exposure at equivalent probing depths of 0.6, 1 and 2 nm, respectively. The orange peaks correspond to the (b) Fe environment (c) Sn environment of pristine FeSn. The oxide peaks are depicted in blue and the green doublet seen in the Sn 3d spectra is signal from the Sn in kagome layer.

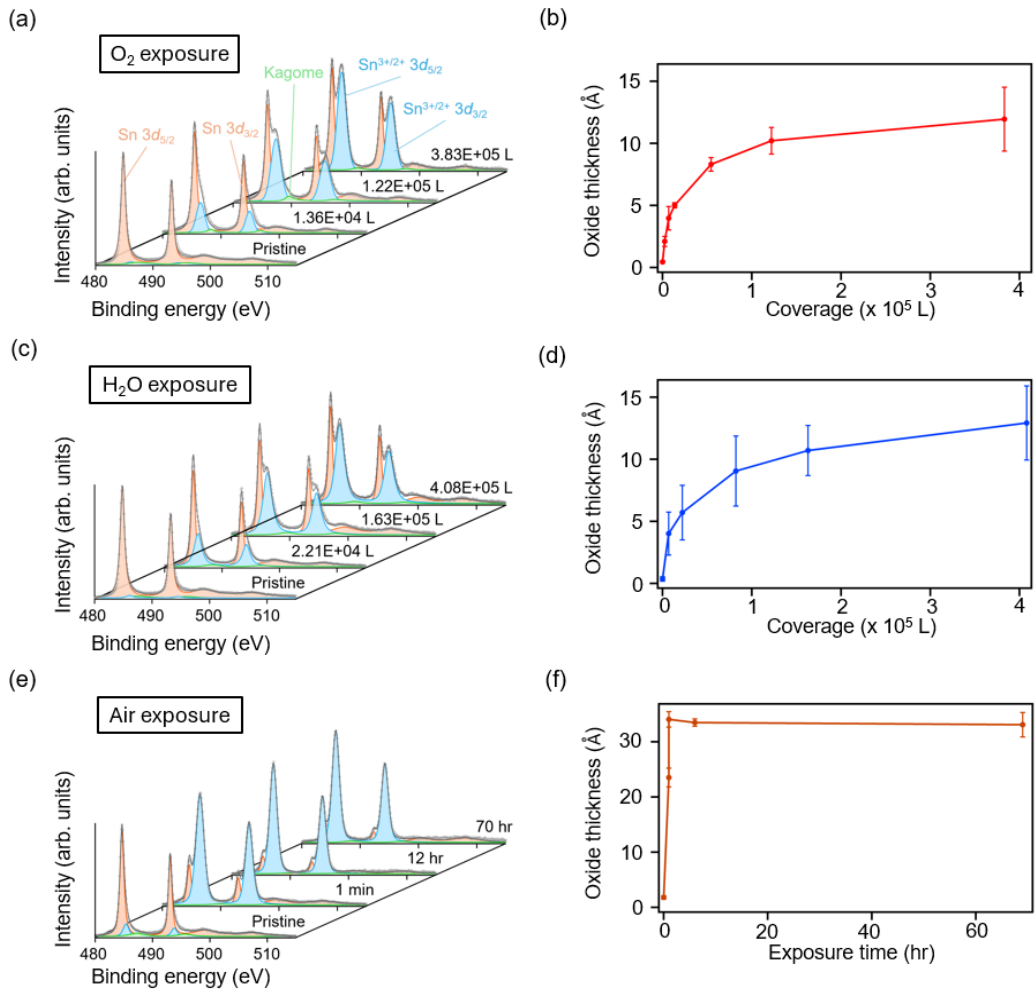

Figure 5. Comparison between $O_2$, $H_2O$ and air exposure and the corresponding oxide thickness. The evolution of the Sn 3d core level for (a) $O_2$ exposure (c) $H_2O$ exposure and (e) air exposure, is shown as a function of coverage (in units of Langmuir) or duration (in units of hours), all taken at an incident photon energy of 1340 eV. The corresponding thicknesses of the oxide layer are shown in figures (b), (d) and (f) respectively calculated as described in main text.

# Supplementary Information

## Reactivity of ultra-thin Kagome Metal FeSn towards Oxygen and Water

James Blyth[1,4,⊥], Sadhana Sridhar[1,⊥], Mengting Zhao[1,3,4,⊥,*], Sajid Ali[2,4], Thi Hai Yen Vu[1,4], Qile Li[1,4], Johnathon Maniatis[1], Grace Causer[1,4], Michael S. Fuhrer[1,4] Nikhil V. Medhekar[2,4], Anton Tadich[3,4], Mark Edmonds[1,4,5]

[1] School of Physics and Astronomy, Monash University, Clayton, VIC 3800 Australia

[2] School of Materials Engineering, Monash University, Clayton, VIC 3800 Australia

[3] Australian Synchrotron, Clayton, VIC 3168 Australia

[4] ARC Centre for Future Low Energy Electronics Technologies, Monash University, Clayton, VIC 3800, Australia

[5] ANFF-VIC Technology Fellow, Melbourne Centre for Nanofabrication, Victorian Node of the Australian National Fabrication Facility, Clayton, VIC 3168, Australia

⊥ These authors contributed equally to this work

* Corresponding author: mengting.zhao@monash.edu

## Table of Contents

## 1. Topographic information of FeSn films with various thickness

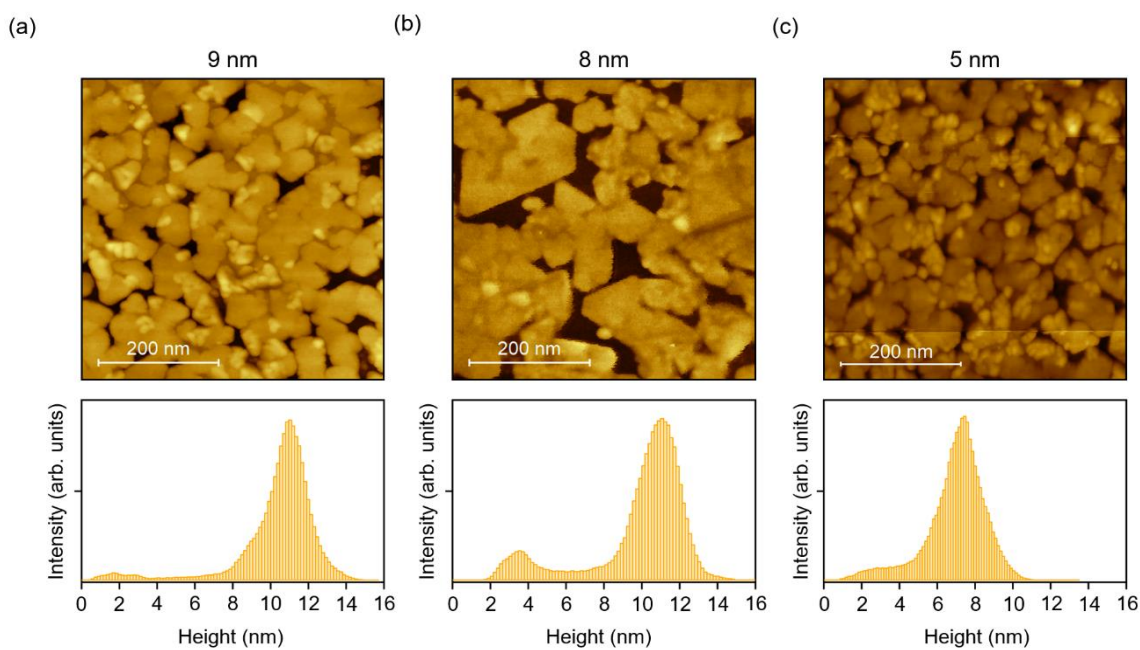

Figure S1. Large-area FeSn film quality was demonstrated using atomic force microscopy (AFM). (a)(b)(c) top panel: 500 nm x 500 nm AFM image taken in contact mode of three FeSn films. Bottom panel: Height profile histogram showing the average film thickness is approximately 9 nm, 8 nm and 5 nm respectively.

## 2. Work function of pristine FeSn films

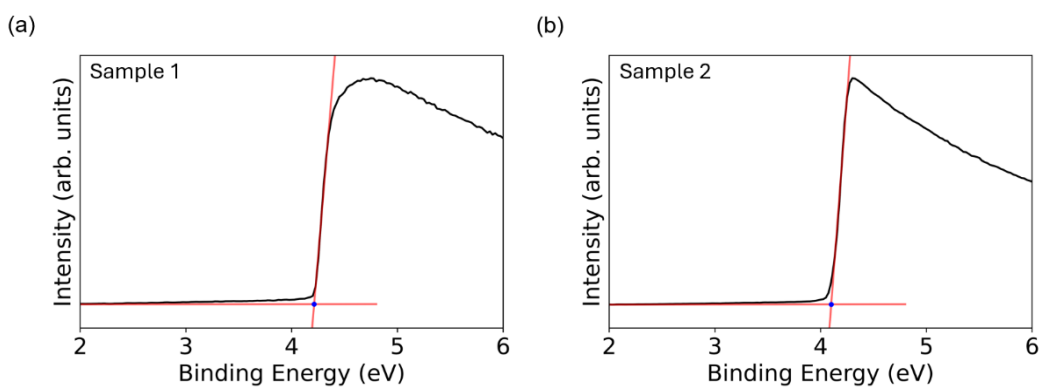

Figure S2 Work function of the two pristine FeSn thin films extracted from the secondary electron cut-off (SECO) measurements (raw data in black). The work function was estimated as the intersecting point (blue) between the background linear fit and cut-off linear fit (both in red) (a)(b) show the pristine the pristine measurements for the FeSn films before exposure to oxygen (a) and water (b). The work functions are $\phi = 4.2\ eV$ and $\phi = 4.1\ eV$, respectively.

## 3. Persistence of Sn 3d Sn-kagome doublet with oxygen exposure

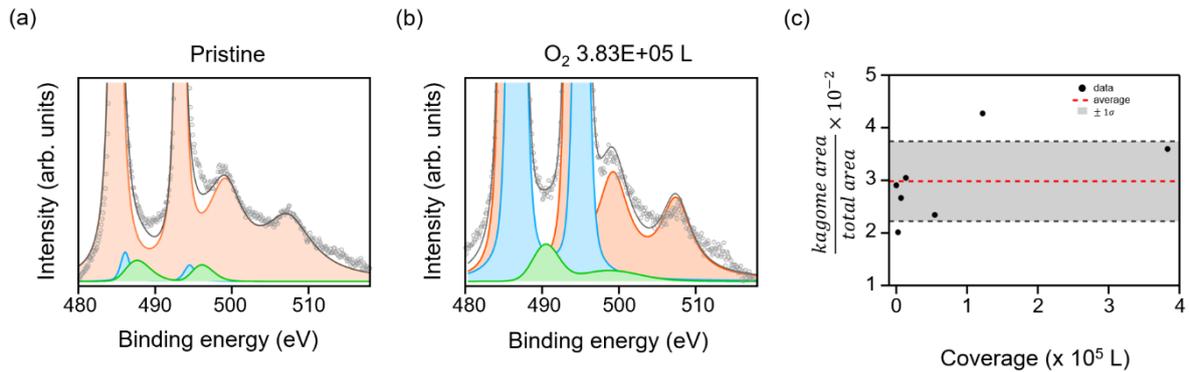

Figure S3. Sn $3d$ core level taken at h$v$ = 1340 eV, demonstrating the preservation of the Sn-kagome doublet. (a) close-up Sn $3d$ spectrum of pristine FeSn. The same color coding from Figure 2 of the main text has been used: orange peaks represent pristine chemical environment, blue represents oxide environments and the green doublet in the Sn $3d$ signal from Sn in the kagome layers. (b) close-up Sn $3d$ spectrum of FeSn after oxygen exposure of $3.83 \times 10^5$ L. (c) The relative proportion of Sn-kagome doublet area over the total Sn $3d$ area versus coverage. The red-dashed line is the average ($3.0 \times 10^{-2}$) and the grey-shaded area is ± one standard deviation ($0.8 \times 10^{-2}$).

## 4. Comparison between oxygen and water exposure.

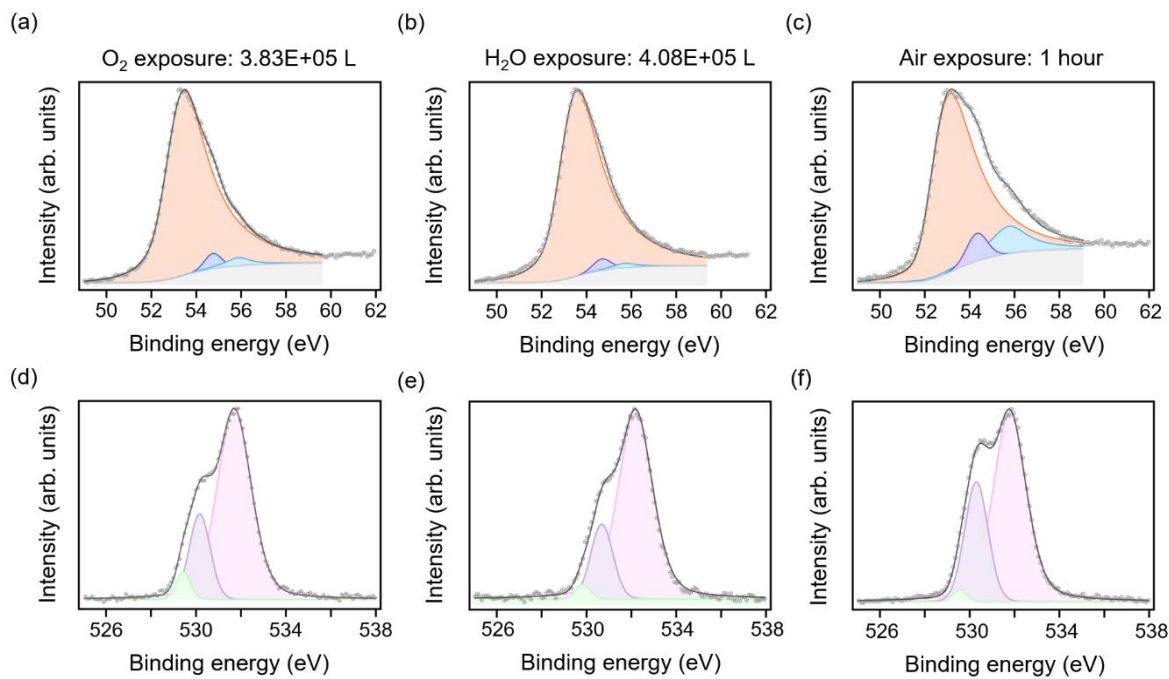

Figure S4. The Fe 3p (at 880 eV) and O1s (at 1340 eV) core levels of each exposure environment highlighting the similarities that indicate identical oxidation mechanisms. The Fe 3p peak is comprised of the pristine, orange peak and two (blue) oxide peaks. The O1s peak has three peaks corresponding to Fe oxides (green), SnO (purple) and $SnO_2$ (pink). (a)(d) are the core levels for the final $O_2$ exposure of $3.83 \times 10^5$ L. (b)(d) are the core levels for the final $H_2O$ exposure of $4.08 \times 10^5$ L. (c) and (f) are the core levels measured after 1 hour of air exposure.